\documentclass[nonacm]{acmart}

\usepackage[utf8]{inputenc}
\usepackage{array,ragged2e}

\usepackage{todonotes}

\newcolumntype{L}[1]{>{\RaggedRight}p{#1}}  
\newcolumntype{R}[1]{>{\RaggedLeft}p{#1}}   
\newcolumntype{C}[1]{>{\Centering}p{#1}}    
\newcolumntype{M}[1]{>{\RaggedRight}m{#1}}  
\newcolumntype{B}[1]{>{\RaggedRight}b{#1}}  

\AtBeginDocument{
  \providecommand\BibTeX{{
    \normalfont B\kern-0.5em{\scshape i\kern-0.25em b}\kern-0.8em\TeX}}}

\acmBooktitle{arXiv}

\begin{document}

\title[Survey about social engineering and the Varni na internetu awareness campaign]{Survey about social engineering and the Varni na internetu awareness campaign, 2020}

\author{Simon Vrhovec}
\email{simon.vrhovec@um.si}
\affiliation{
  \institution{University of Maribor}
  \country{Slovenia}
}

\begin{abstract}
This paper reports on a study aiming to explore factors associated with behavioral intention to follow a social engineering awareness campaign. The objectives of this study were to determine how perceived severity, perceived vulnerability, perceived threat, fear, subjective norm, attitude towards behavior, perceived behavioral control, self-efficacy, response efficacy, trust in authorities, perceived regulation, authorities performance, information sensitivity and privacy concern are associated with individuals' behavioral intention to follow a social engineering awareness campaign. The study employed a cross-sectional research design. A survey was conducted among individuals in Slovenia between January and June 2020. A total of 553 respondents completed the survey providing for N=542 useful responses after excluding poorly completed responses (27.9 percent response rate). The survey questionnaire was developed in English. A Slovenian translation of the survey questionnaire is available.
\end{abstract}


\maketitle

\section{Introduction}

The survey questionnaire was designed to measure theoretical constructs included in the research model. All items were taken or adapted from existing literature to fit the study's context. Table \ref{table:constructs} presents the theoretical constructs included in the research model, their definition in this research, and sources from which construct items were taken or adapted.

\clearpage

\begin{table}[h!]
\caption{\label{table:constructs}Theoretical constructs in the survey questionnaire.}
\small
\begin{tabular}{L{.22\textwidth}L{.63\textwidth}L{.08\textwidth}}
\toprule
Theoretical construct & Definition in this research & Sources \\
\midrule
Perceived severity & The perceived extent of consequences of a successful social engineering attack. & \cite{Fujs2019,Moody2018} \\
Perceived vulnerability & The perceived probability of a successful social engineering attack. & \cite{Fujs2019,Jansen2018} \\
Perceived threat & The perceived extent of threats to the individual posed by social engineering attacks. & \cite{Fujs2019,Liang2010} \\
Fear & The level of the individual’s fear of social engineering. & \cite{Osman1994,Jansen2018} \\
Subjective norm & The perception of social approval from important others regarding following awareness campaign materials. & \cite{Venkatesh2003,Park2007} \\
Attitude towards behavior & An individual's positive versus negative evaluations of following awareness campaign materials. & \cite{Venkatesh2003,Moody2018,Park2007} \\
Perceived behavioral control & The perception of the ease or difficulty of following awareness campaign materials. & \cite{Park2007,Venkatesh2003} \\
Self-efficacy & The individual’s self-efficacy when implementing social engineering countermeasures (i.e., following awareness campaign materials). & \cite{Johnston2010,Anderson2010} \\
Response efficacy & The perceived efficacy of social engineering countermeasures (i.e., following awareness campaign materials). & \cite{Moody2018,Jansen2018} \\
Trust in authorities & The degree of trust (trusting beliefs) in state authorities. & \cite{McKnight2002,Jansen2018} \\
Perceived regulation & The perceived appropriateness of the regulative framework for fighting social engineering. & \cite{Fujs2019} \\
Authorities performance & The perceived performance of state authorities when dealing with social engineering. & \cite{Crow2017} \\
Information sensitivity & The perceived sensitivity of an individual's online information. &  \cite{Fujs2019} \\
Privacy concern & The extent of concerns regarding privacy online. & \cite{Fujs2019} \\
Behavioral intention & The level of individual’s motivation to follow awareness campaign materials in the near future. & \cite{Park2007} \\
\bottomrule
\end{tabular}
\end{table}

\section{Method}

\subsection{Survey instrument}

To test the research model a survey questionnaire was developed. Adapted and new questionnaire items (i.e., perceived severity, perceived vulnerability, perceived threat, fear, subjective norm, attitude towards behavior, perceived behavioral control, self-efficacy, response efficacy, trust in authorities, perceived regulation, authorities performance, information sensitivity, and behavioral intention) were developed by following a predefined protocol.

The questionnaire was first developed in English and then translated into Slovenian by a translator. The Slovenian questionnaire has been pre-tested by 3 independent respondents who provided feedback on its clarity. Based on the received feedback, the Slovenian questionnaire was reviewed to remove any ambiguity. Items were reworded, added, and deleted in the pre-test. To ensure the consistency between the Slovenian and English questionnaire, the Slovenian questionnaire was translated back to English. No significant differences in the meaning between the original items in English and back-translations were noticed. The English questionnaire was however reviewed to update the items and to remove any ambiguity based on the back-translation.

Table \ref{table:questionnaire_en} presents the survey questionnaire in English and Table \ref{table:questionnaire_si} presents the Slovenian translation of the survey questionnaire. All items were measured using a 5-point Likert scale as presented in Table \ref{table:likert}.

\clearpage

\begin{table}[h!]
\caption{\label{table:questionnaire_en}Survey questionnaire items (English original).}
\footnotesize
\begin{tabular}{L{.24\textwidth}L{.71\textwidth}}
\toprule
Construct & Prompt/Item \\
\midrule
Perceived severity (PS) & Mark your agreement with the statements about social engineering: \\
 & PS1. My personal data acquired with social engineering could be misused for criminal purposes. \\
 & PS2. My personal data collected with social engineering could be misused against me. \\
 & PS3. Stealing of my personal data with social engineering would be a serious problem for me. \\

Perceived vulnerability (PV) & Mark your agreement with the statements about social engineering: \\
 & PV1. I am very vulnerable to social engineering. \\
 & PV2. I can easily become a victim of social engineering. \\
 & PV3. It is likely that I will become a victim of social engineering in the near future. \\

Perceived threat (PT) & Mark your agreement with the statements about social engineering: \\
 & PT1. I feel threatened by social engineering. \\
 & PT2. Social engineering threatens me. \\
 & PT3. Social engineering is a danger to me. \\

Fear (F) & Mark your agreement with the statements about social engineering: \\
 & F1. I am afraid of social engineering. \\
 & F2. Social engineering is very frightening. \\
 & F3. I am afraid of being victimized by social engineering. \\

Subjective norm (SN) & Mark your agreement with the statements about following the \textit{Varni na internetu} awareness campaign: \\
 & SN1. People who are important to me think that I should follow awareness campaign materials. \\
 & SN2. Most people whose opinion I value consider that I should follow awareness campaign materials. \\
 & SN3. It is expected of me that I follow awareness campaign materials. \\

Attitude towards behavior (AtB) & Mark your agreement with the statements about following the \textit{Varni na internetu} awareness campaign: \\
 & AtB1. Following awareness campaign materials is a very good idea. \\
 & AtB2. Following awareness campaign materials would be very wise. \\
 & AtB3. Following awareness campaign materials is very beneficial. \\

Perceived behavioral control (PBC) & Mark your agreement with the statements about following the \textit{Varni na internetu} awareness campaign: \\
 & PBC1. If I want to, I can follow awareness campaign materials. \\
 & PBC2. I know how to follow awareness campaign materials. \\
 & PBC3. I have the resources necessary to follow awareness campaign materials. \\

Self-efficacy (SE) & Mark your agreement with the statements about following the \textit{Varni na internetu} awareness campaign: \\
 & SE1. It is easy to follow awareness campaign materials. \\
 & SE2. I would feel comfortable following awareness campaign materials. \\
 & SE3. I am able to follow awareness campaign materials without much effort. \\

Response efficacy (RE) & Mark your agreement with the statements about following the \textit{Varni na internetu} awareness campaign: \\
 & RE1. Following awareness campaign materials lowers the success of social engineering. \\
 & RE2. Following awareness campaign materials helps in preventing social engineering. \\
 & RE3. If I follow awareness campaign materials, I am less likely to be victimized by social engineering. \\

Trust in authorities (TiA) & Mark your agreement with the statements about state authorities (e.g., police, national cybersecurity response center SI-CERT): \\
 & TiA1. I believe that the state authorities would act in my best interest. \\
 & TiA2. I would characterize state authorities as honest. \\
 & TiA3. I trust state authorities. \\

Perceived regulation (PR) & Mark your agreement with the statements about the adequacy of the legislation for fighting social engineering: \\
 & PR1. Our legislation provides for appropriate measures for fighting social engineering. \\
 & PR2. The international legislation provides for appropriate measures for fighting social engineering. \\
 & PR3. The government does enough to fight social engineering. \\

Authorities performance (AP) & Mark your agreement with the statements about the activities of state authorities (e.g., police, national cybersecurity response center SI-CERT) for fighting social engineering: \\
 & AP1. State authorities are successfully dealing with social engineering. \\
 & AP2. State authorities are successfully working with internet users to address social engineering. \\
 & AP3. State authorities are successfully preventing social engineering. \\

Information sensitivity (IS) & Mark your agreement with the statements about your activity online: \\
 & IS1. I consider the content of my e-mails as very sensitive. \\
 & IS2. I consider data on which websites I visit as very sensitive. \\
 & IS3. I consider data on what I do online as very sensitive. \\

Privacy concern (PC) & Mark your agreement with the statements about your personal data online: \\
 & PC1. It highly bothers me when websites ask me about my personal data. \\
 & PC2. I always think twice before submitting my personal data online. \\
 & PC3. I am very concerned that websites collect too much personal data about me. \\

Behavioral intention (BI) & Mark your agreement with the statements about following the \textit{Varni na internetu} awareness campaign: \\
 & BI1. I intend to follow awareness campaign materials in the near future. \\
 & BI2. I have it in my mind to follow awareness campaign materials in the near future. \\
 & BI3. I will follow awareness campaign materials in the near future. \\
\bottomrule
\end{tabular}
\end{table}

\clearpage

\begin{table}[h!]
\caption{\label{table:questionnaire_si}Survey questionnaire items (Slovenian translation).}
\footnotesize
\begin{tabular}{L{.24\textwidth}L{.71\textwidth}}
\toprule
Construct & Prompt/Item \\
\midrule
Perceived severity (PS) & Označite svoje strinjanje z izjavami o družbenem inženiringu: \\
 & PS1. Moji osebni podatki, pridobljeni z družbenim inženiringom, bi bili lahko zlorabljeni v kriminalne namene. \\
 & PS2. Moji osebni podatki, pridobljeni z družbenim inženiringom, bi bili lahko zlorabljeni zoper mene. \\
 & PS3. Kraja mojih osebnih podatkov z družbenim inženiringom bi bila zame resna težava. \\

Perceived vulnerability (PV) & Označite svoje strinjanje z izjavami o družbenem inženiringu: \\
 & PV1. Zelo sem ranljiv za družbeni inženiring. \\
 & PV2. Zlahka lahko postanem žrtev družbenega inženiringa. \\
 & PV3. Verjetno bom postal žrtev družbenega inženiringa v bližnji prihodnosti. \\

Perceived threat (PT) & Označite svoje strinjanje z izjavami o družbenem inženiringu: \\
 & PT1. Zaradi družbenega inženiringa se počutim ogroženega. \\
 & PT2. Družbeni inženiring me ogroža. \\
 & PT3. Družbeni inženiring mi predstavlja nevarnost. \\

Fear (F) & Označite svoje strinjanje z izjavami o družbenem inženiringu: \\
 & F1. Bojim se družbenega inženiringa. \\
 & F2. Družbeni inženiring je zelo zastrašujoč. \\
 & F3. Bojim se, da bi postal žrtev družbenega inženiringa. \\

Subjective norm (SN) & Označite svoje strinjanje z izjavami o spremljanju vsebin programa ozaveščanja \textit{Varni na internetu}: \\
 & SN1. Ljudje, ki so mi pomembni, mislijo, da bi moral spremljati vsebine programa ozaveščanja. \\
 & SN2. Večina ljudi, katerih mnenje cenim, meni, da bi moral spremljati vsebine programa ozaveščanja. \\
 & SN3. Od mene se pričakuje, da spremljam vsebine programa ozaveščanja. \\

Attitude towards behavior (AtB) & Označite svoje strinjanje z izjavami o spremljanju vsebin programa ozaveščanja \textit{Varni na internetu}: \\
 & AtB1. Spremljanje vsebin programa ozaveščanja je zelo dobra ideja. \\
 & AtB2. Spremljanje vsebin programa ozaveščanja bi bilo zelo pametno. \\
 & AtB3. Spremljanje vsebin programa ozaveščanja je zelo koristno. \\

Perceived behavioral control (PBC) & Označite svoje strinjanje z izjavami o spremljanju vsebin programa ozaveščanja \textit{Varni na internetu}: \\
 & PBC1. Če želim, lahko spremljam vsebine programa ozaveščanja. \\
 & PBC2. Vem, kako spremljati vsebine programa ozaveščanja. \\
 & PBC3. Imam vse potrebno za spremljanje vsebin programa ozaveščanja. \\

Self-efficacy (SE) & Označite svoje strinjanje z izjavami o spremljanju vsebin programa ozaveščanja \textit{Varni na internetu}: \\
 & SE1. Vsebine programa ozaveščanja je preprosto spremljati. \\
 & SE2. Brez težav bi lahko spremljal vsebine programa ozaveščanja. \\
 & SE3. Sposoben sem spremljati vsebine programa ozaveščanja brez posebnega napora. \\

Response efficacy (RE) & Označite svoje strinjanje z izjavami o spremljanju vsebin programa ozaveščanja \textit{Varni na internetu}: \\
 & RE1. Spremljanje vsebin programa ozaveščanja znižuje uspešnost družbenega inženiringa. \\
 & RE2. Spremljanje vsebin programa ozaveščanja pomaga pri preprečevanju družbenega inženiringa. \\
 & RE3. Če spremljam vsebine programa ozaveščanja, je manj verjetno, da postanem žrtev družbenega inženiringa. \\

Trust in authorities (TiA) & Označite svoje strinjanje z izjavami o državnih organih (npr. policija, nacionalni odzivni center za kibernetsko varnost SI-CERT): \\
 & TiA1. Verjamem, da državni organi delujejo v mojem najboljšem interesu. \\
 & TiA2. Državne organe bi opredelil kot poštene. \\
 & TiA3. Zaupam državnim organom. \\

Perceived regulation (PR) & Označite svoje strinjanje z izjavami o ustreznosti pravne ureditve za boj proti družbenemu inženiringu: \\
 & PR1. Naša zakonodaja predvideva ustrezne ukrepe za boj proti družbenemu inženiringu. \\
 & PR2. Mednarodna zakonodaja predvideva ustrezne ukrepe za boj proti družbenemu inženiringu. \\
 & PR3. Država stori dovolj za boj proti družbenemu inženiringu. \\

Authorities performance (AP) & Označite svoje strinjanje z izjavami o aktivnosti državnih organov (npr. policija, nacionalni odzivni center za kibernetsko varnost SI-CERT) pri soočanju z družbenim inženiringom: \\
 & AP1. Državni organi se uspešno spopadajo z družbenim inženiringom. \\
 & AP2. Državni organi uspešno sodelujejo z uporabniki interneta pri reševanju družbenega inženiringa. \\
 & AP3. Državni organi uspešno preprečujejo družbeni inženiring. \\

Information sensitivity (IS) & Označite svoje strinjanje z izjavami o vaši aktivnosti na spletu: \\
 & IS1. Vsebino svoje elektronske pošte dojemam kot zelo občutljivo. \\
 & IS2. Podatke o tem, katere spletne strani obiskujem, dojemam kot zelo občutljive. \\
 & IS3. Podatke o tem, kaj delam na spletu, dojemam kot zelo občutljive. \\

Privacy concern (PC) & Označite svoje strinjanje z izjavami o vaših osebnih podatkih na spletu: \\
 & PC1. Zelo me moti, ko me spletne strani sprašujejo po osebnih podatkih. \\
 & PC2. Preden posredujem svoje osebne podatke preko spleta, vedno premislim dvakrat. \\
 & PC3. Zelo me skrbi, da spletne strani o meni zbirajo preveč osebnih podatkov. \\

Behavioral intention (BI) & Označite svoje strinjanje z izjavami o spremljanju vsebin programa ozaveščanja \textit{Varni na internetu}: \\
 & BI1. V bližnji prihodnosti nameravam spremljati vsebine programa ozaveščanja. \\
 & BI2. V mislih imam spremljanje vsebin programa ozaveščanja v bližnji prihodnosti. \\
 & BI3. V bližnji prihodnosti bom spremljal vsebine programa ozaveščanja. \\
\bottomrule
\end{tabular}
\end{table}

\clearpage

\begin{table}[h!]
\caption{\label{table:likert}5-point Likert scale.}
\small
\begin{tabular}{L{.06\textwidth}L{.19\textwidth}L{.19\textwidth}}
\toprule
Score & English & Slovenian \\
\midrule
1 & Strongly disagree & Močno se ne strinjam \\
2 & Disagree & Se ne strinjam \\
3 & Neutral & Nevtralno \\
4 & Agree & Se strinjam \\
5 & Strongly agree & Močno se strinjam \\
\bottomrule
\end{tabular}
\end{table}

\subsection{Data collection}

We conducted the survey with the Slovenian translation of the questionnaire among individuals in Slovenia who were at least 15 years old between 6 January 2020 and 24 June 2020. Respondents were recruited through University of Maribor students who were asked to distribute the survey questionnaire to their family and friends. The students were not compensated for distributing the survey questionnaire. Also, the respondents did not receive any compensation for taking the survey. All batches of questionnaires were checked for any signs of misconduct on the part of the students (e.g., asking questions related to respondents upon returning the questionnaires, checking if the same pen was used, style of writing, face similarity of answers, logical errors). No signs of misconduct were noticed. A total of 553 questionnaires were returned. After excluding poorly completed responses (responses with over 50 percent of missing values or standard deviation equal to 0 for constructs perceived severity, perceived vulnerability, perceived threat, fear, subjective norm, attitude towards behavior, perceived behavioral control, self-efficacy, response efficacy, trust in authorities, perceived regulation, authorities performance, information sensitivity, and behavioral intention), we were left with 542 useful responses providing for a response rate of 27.9 percent as presented in Table \ref{table:sample}.

\begin{table}[h!]
\caption{\label{table:sample}Sample with the number of distributed questionnaires, number of responses, and number of useful responses ($N$) after excluding poorly completed responses.}
\small
\begin{tabular}{L{.02\textwidth}L{.18\textwidth}R{.21\textwidth}R{.09\textwidth}R{.04\textwidth}}
\toprule
ID & Name & Distributed questionnaires & Responses & $N$ \\
\midrule
1 & Individuals in Slovenia & 1982 & 553 & 542 \\
\bottomrule
\end{tabular}
\end{table}

Due to the sensitive nature of the survey topic, safeguards were put in place to encourage participation and honest responses. First, the respondents were informed about the voluntariness and anonymity of participating in the survey. Next, the respondents were assured that the collected data will be used for research purposes only. No special incentives were offered to encourage participation in the survey.

The first page of the survey is presented in Table \ref{table:first-page}.

\clearpage

\begin{table}[h!]
\caption{\label{table:first-page}The first page of the survey.}
\small
\begin{tabular}{L{.47\textwidth}L{.47\textwidth}}
\toprule
English original & Slovenian translation \\
\midrule
\textit{Study on social engineering and the Varni na internetu awareness campaign} \newline ~ \newline Dear Sirs! \newline ~ \newline You are cordially invited to participate in a study on social engineering and the Varni na internetu awareness campaign conducted by the University of Maribor. Estimated time to complete the survey is 6-10 minutes. \newline ~ \newline Social engineering refers to the use of deception in order to manipulate individuals to disclose confidential or personal information that may then be used for criminal purposes. \newline ~ \newline Don’t worry if you’re not sure what social engineering and the Varni na internetu awareness campaign are. A short presentation of the most problematic types of social engineering in the last year, and a presentation of the Varni na internetu awareness campaign are part of the survey. \newline ~ \newline Participation in the research is voluntary and anonymous. Data will be used exclusively for research purposes. \newline ~ \newline Contact person: \newline Simon Vrhovec <simon.vrhovec@um.si> \newline ~ \newline Kind regards, \newline Simon Vrhovec & \textit{Raziskava o družbenem inženiringu in programu ozaveščanja Varni na internetu} \newline ~ \newline Spoštovani! \newline ~ \newline Vljudno vabljeni k sodelovanju v raziskavi o družbenem inženiringu in programu ozaveščanja Varni na internetu, ki jo izvajamo na Univerzi v Mariboru. Predviden čas izpolnjevanja ankete je 6-10 minut. \newline ~ \newline Družbeni inženiring (angl. social engineering) se nanaša na uporabo zavajanja za manipulacijo posameznikov, da razkrijejo zaupne ali osebne podatke, ki se lahko nato uporabijo v kriminalne namene. \newline ~ \newline Naj vas ne skrbi, če niste prepričani, kaj sta družbeni inženiring in program ozaveščanja Varni na internetu. Kratka predstavitev najbolj problematičnih vrst družbenega inženiringa v zadnjem letu in predstavitev programa ozaveščanja Varni na internetu sta namreč del ankete. \newline ~ \newline Sodelovanje v raziskavi je prostovoljno in anonimno, podatki pa bodo uporabljeni izključno za raziskovalne namene. \newline ~ \newline Kontaktna oseba: \newline Simon Vrhovec <simon.vrhovec@um.si> \newline ~ \newline Z lepimi pozdravi, \newline Simon Vrhovec \\
\bottomrule
\end{tabular}
\end{table}

\section{Results}

\subsection{Sample}

Demographic characteristics of the sample are presented in Table \ref{table:demographics}.

\begin{table}[h!]
\caption{\label{table:demographics}Demographic characteristics of the sample.}
\footnotesize
\begin{tabular}{L{.18\textwidth}L{.24\textwidth}R{.08\textwidth}}
\toprule
Characteristic & Value & Frequency \\
\midrule
Gender & 1 -- Male & 225 \\
 & 2 -- Female & 315 \\
 & \textit{Missing} & 13 \\
 
Formal education & 1 -- High school or less & 348 \\
 & 2 -- Bachelor's degree & 139 \\
 & 3 -- Master's degree & 48 \\
 & 4 -- PhD degree & 5 \\
 & \textit{Missing} & 13 \\
 
Employment status & 1 -- Student & 267 \\
 & 2 -- Employed & 240 \\
 & 3 -- Unemployed & 11 \\
 & 4 -- Retired & 20 \\
 & \textit{Missing} & 15 \\

Age & 15 & 1 \\
 & 16 & 3 \\
 & 17 & 1 \\
 & 18 & 7 \\
 & 19 & 47 \\
 & 20 & 71 \\
 & 21 & 67 \\
 & 22 & 31 \\
 & 23 & 28 \\
 & 24 & 18 \\
 & 25 & 18 \\
 & 26 & 15 \\
 & 27 & 9 \\
 & 28 & 8 \\
 & 29 & 10 \\
 & 30 & 19 \\
 & 31 & 4 \\
 & 32 & 10 \\
 & 33 & 7 \\
 & 34 & 3 \\
 & 35 & 10 \\
 & 36 & 3 \\
 & 37 & 3 \\
 & 38 & 7 \\
 & 39 & 7 \\
 & 40 & 5 \\
 & 41 & 4 \\
 & 42 & 8 \\
 & 43 & 8 \\
 & 44 & 8 \\
 & 45 & 8 \\
 & 46 & 4 \\
 & 47 & 3 \\
 & 48 & 9 \\
 & 49 & 7 \\
 & 50 & 10 \\
 & 51 & 5 \\
 & 52 & 8 \\
 & 53 & 4 \\
 & 54 & 7 \\
 & 55 & 6 \\
 & 56 & 2 \\
 & 57 & 4 \\
 & 59 & 2 \\
 & 60 & 5 \\
 & 61 & 3 \\
 & 62 & 3 \\
 & 64 & 1 \\
 & 65 & 2 \\
 & 66 & 2 \\
 & 68 & 1 \\
 & 69 & 1 \\
 & 73 & 1 \\
 & 74 & 1 \\
 & 75 & 1 \\
 & 78 & 1 \\
 & \textit{Missing} & 12 \\
\bottomrule
\end{tabular}
\end{table}

\clearpage

\subsection{Frequencies}

Frequencies of all variables for measured theoretical constructs are presented in Table \ref{table:frequencies}.

\begin{table}[h!]
\caption{\label{table:frequencies}Frequencies of variables.}
\small
\begin{tabular}{L{.07\textwidth}R{.03\textwidth}R{.03\textwidth}R{.03\textwidth}R{.03\textwidth}R{.03\textwidth}R{.07\textwidth}R{.07\textwidth}R{.07\textwidth}}
\toprule
Variable & 1 & 2 & 3 & 4 & 5 & Valid & Missing & Total \\
\midrule
IS1 & 14 & 93 & 200 & 176 & 69 & 552 & 1 & 553 \\
IS2 & 12 & 98 & 208 & 169 & 64 & 551 & 2 & 553 \\
IS3 & 10 & 85 & 153 & 201 & 97 & 546 & 7 & 553 \\

PC1 & 3 & 19 & 50 & 244 & 236 & 552 & 1 & 553 \\
PC2 & 2 & 26 & 51 & 201 & 271 & 551 & 2 & 553 \\
PC3 & 3 & 27 & 88 & 196 & 237 & 551 & 2 & 553 \\

TiA1 & 19 & 82 & 236 & 186 & 30 & 553 & 0 & 553 \\
TiA2 & 29 & 103 & 236 & 153 & 29 & 550 & 3 & 553 \\
TiA3 & 35 & 103 & 240 & 142 & 30 & 550 & 3 & 553 \\

AP1 & 13 & 116 & 303 & 103 & 3 & 538 & 15 & 553 \\
AP2 & 14 & 104 & 294 & 117 & 10 & 539 & 14 & 553 \\
AP3 & 17 & 146 & 297 & 72 & 3 & 535 & 18 & 553 \\

PR1 & 8 & 128 & 252 & 146 & 3 & 537 & 16 & 553 \\
PR2 & 10 & 80 & 270 & 164 & 12 & 536 & 17 & 553 \\
PR3 & 26 & 211 & 231 & 62 & 7 & 537 & 16 & 553 \\

PV1 & 47 & 162 & 185 & 125 & 15 & 534 & 19 & 553 \\
PV2 & 49 & 182 & 115 & 162 & 29 & 537 & 16 & 553 \\
PV3 & 84 & 187 & 183 & 73 & 10 & 537 & 16 & 553 \\

PS1 & 24 & 94 & 129 & 248 & 42 & 537 & 16 & 553 \\
PS2 & 19 & 70 & 116 & 270 & 63 & 538 & 15 & 553 \\
PS3 & 13 & 52 & 118 & 219 & 136 & 538 & 15 & 553 \\

PT1 & 18 & 147 & 234 & 127 & 14 & 540 & 13 & 553 \\
PT2 & 27 & 159 & 253 & 93 & 8 & 540 & 13 & 553 \\
PT3 & 22 & 120 & 244 & 135 & 18 & 539 & 14 & 553 \\

F1 & 37 & 165 & 203 & 111 & 23 & 539 & 14 & 553 \\
F2 & 38 & 125 & 193 & 144 & 38 & 538 & 15 & 553 \\
F3 & 39 & 143 & 194 & 129 & 33 & 538 & 15 & 553 \\

SE1 & 8 & 55 & 177 & 269 & 28 & 537 & 16 & 553 \\
SE2 & 6 & 46 & 138 & 317 & 33 & 540 & 13 & 553 \\
SE3 & 5 & 32 & 136 & 299 & 63 & 535 & 18 & 553 \\

RE1 & 3 & 36 & 191 & 262 & 45 & 537 & 16 & 553 \\
RE2 & 6 & 20 & 167 & 303 & 43 & 539 & 14 & 553 \\
RE3 & 2 & 36 & 123 & 292 & 83 & 536 & 17 & 553 \\

PBC1 & 1 & 9 & 89 & 363 & 76 & 538 & 15 & 553 \\
PBC2 & 8 & 59 & 187 & 242 & 43 & 539 & 14 & 553 \\
PBC3 & 10 & 45 & 168 & 250 & 66 & 539 & 14 & 553 \\

SN1 & 23 & 118 & 223 & 152 & 22 & 538 & 15 & 553 \\
SN2 & 28 & 100 & 239 & 150 & 20 & 537 & 16 & 553 \\
SN3 & 33 & 105 & 223 & 157 & 20 & 538 & 15 & 553 \\

AtB1 & 3 & 12 & 68 & 336 & 120 & 539 & 14 & 553 \\
AtB2 & 2 & 6 & 67 & 321 & 143 & 539 & 14 & 553 \\
AtB3 & 2 & 4 & 72 & 309 & 151 & 538 & 15 & 553 \\

BI1 & 21 & 73 & 206 & 208 & 32 & 540 & 13 & 553 \\
BI2 & 21 & 74 & 181 & 224 & 38 & 538 & 15 & 553 \\
BI3 & 24 & 63 & 204 & 206 & 41 & 538 & 15 & 553 \\
\bottomrule
\end{tabular}
\end{table}

\section{Discussion}

This paper presents the results of a survey about social engineering and the \textit{Varni na internetu} awareness campaign. Future studies may focus on other factors associated with individuals' motivation to follow awareness campaign materials. Such studies would be beneficial both to better explain the motivation of internet users to follow awareness campaigns, and to better understand the associations between different factors associated with it.

\section*{Acknowledgements}

We would like to express our sincere gratitude to the respondents who took their time to participate in our survey. We would also like to thank the students who helped with data collection, and Luka Jelovčan for data screening and entry.

This study is part of a research project \textit{Safety and security of cyberspace users -- Criminological, victimological and preventative aspects} (J5-9345, 2018-2020) funded by the Slovenian Research Agency, and carried out by the University of Maribor.

\bibliographystyle{ACM-Reference-Format}
\bibliography{main}

\end{document}